\documentstyle{article}

\newcommand{\be}{\begin{equation}}
\newcommand{\ee}{\end{equation}}
\newcommand{\beq}{\begin{eqnarray}}
\newcommand{\eeq}{\end{eqnarray}}
\newcommand{\bed}{\begin{displaymath}}
\newcommand{\eed}{\end{displaymath}}
\begin{document}
\thispagestyle{empty}
\begin{center}
\vspace{1.0in}

{\Large\bf{Hamilton-Jacobi approach to pre-big bang cosmology and the 
       problem of initial conditions}}

\vspace{0.5in}
 
\vspace{0.5in}

Piret Kuusk$^{1}$ and  Margus Saal$^{2}$\\

\end{center}

\vspace{1in}

\begin{abstract}
The Hamilton-Jacobi equation for the string cosmology
is solved using the gradient expansion method. 
The zeroth order solution is taken to be the standard pre-big bang model  
and the second order solution is found for the dilaton and the three-metric.
It indicates  that corrections generated by inhomogeneities of 
the seed metric are suppressed near
the singularity and are growing towards the asymptotic past, but 
corrections generated by the dilaton inhomogeneities  
are growing near the singularity
and are suppressed in the past. Possible influences of initial 
metric inhomogeneities on the pre-big bang superinflation are discussed. 


\end{abstract}

\vspace{0.5in}

$^1$ Institute of Physics, University of Tartu,
Riia 142, Tartu 51014, Estonia. 

$^2$ University of Tartu, T\"ahe 4, Tartu 51010, Estonia. \\ 
    E-mail: margus@hexagon.fi.tartu.ee  

\newpage


\section {Introduction} 

Observations show that at present our Universe is rather homogeneous and
isotropic.  It is commonly believed that the present homogeneity
was achieved from a chaotic (inhomogeneous) initial state by inflation.
In the  string cosmology \cite{homepage, lidsey},
the same can be obtained by a pre-big bang superinflation.
The pre-big bang era describes a possible 
evolution of the Universe from the string perturbative vacuum 
where classical equations for gravitational and moduli fields hold 
until the curvature and coupling reach their maximum and both, string 
and quantum corrections, become crucial. It is believed (but not yet proved)
that then a  graceful exit transition 
\cite{gaspe3, WdW} to the usual 
Friedmann-Robertson-Walker Universe occurs, and standard cosmological 
picture, with some shades, works well. 

Besides the graceful exit problem, there is the problem of initial conditions:
it has been argued that
pre-big bang initial conditions have to be fine-tuned in order to
give expected results \cite{coule}.
Turner and Weinberg \cite{turnwein} concluded  that curvature terms 
postpone the onset of inflation and can prevent getting sufficient amount 
of inflation before higher-order loop and string corrections become important.
Kaloper, Linde and Bousso \cite{kaloper1} have argued that horizon and 
flatness problems will be solved if the Universe at the onset of 
inflation is exponentially large and homogeneous.
Clancy et al \cite{clancy} have  found, that the constraints 
for sufficient amount
of inflation in anisotropic models are stronger than in the isotropic case.
Numerical calculations presented by Maharana et al \cite{maharana} 
and Chiba \cite{chiba1}   
contain controversial results concerning the decay of 
initial inhomogeneities. 
Counterarguments for justifying the pre-big bang inflationary model 
have been given in \cite{vene1, maggiore1, buonanno1, gasperini6}.      

There are several papers which discuss the role of initial 
inhomogeneities  in  cosmological models (for a review see \cite{gold}).  
Using the gradient expansion method 
developed by Salopek et al \cite{salopek, perry}, evolution of 
inhomogeneities in 
cosmological models which contain cosmological constant and radiation field
\cite{ishihara1},
massive minimally coupled scalar field  \cite{ishihara2}, 
or perfect fluid \cite{chiba2} were considered.
The Brans-Dicke theory with a dust and cosmological constant was investigated 
by Soda et al \cite{soda} using the 
gradient expansion of the corresponding Hamilton-Jacobi equation.     
The direct method to solve the Einstein equations expanded in 
spatial gradients
is  developed and applied to various models \cite{comer}.
The closest to our approach is the paper by Nambu and Taruya \cite{nambu}
where the role of initial inhomogeneities in the conventional inflationary
cosmology is discussed in the framework of long-wavelength approximation.

In the string cosmology,  inhomogeneous  
spherically and cylindrically 
symmetric models are investigated by Barrow et al \cite{barrow1, barrow2}
and Feinstein et al \cite{feinstein} using methods familiar in the
general relativity. Buonanno et al \cite{damour} link the cosmological
scenario with collapsing initial gravitational waves.

In the following we shall  use the gradient expansion method 
developed by Saygili \cite{saygili}
 for investigating the effect of small 
inhomogeneities and related three-curvature. 
We assume the standard pre-big bang cosmology  to be valid as the
zeroth order approximation and include  inhomogeneities
as  small perturbations. They generate  small  three-curvature  
which may affect the onset and duration of superinflation.   
Although the long-wavelength approximation is valid in a region
where spatial derivatives are negligible and also  inhomogeneities 
must be constrained, 
it can be used for investigating the evolution trends of
small inhomogeneities and deriving combined constraints
on their size and the duration of inflation.
We demonstrate that the initial spatial curvature is suppressed during the
superinflation and therefore neglecting the spatial gradients improves
in time. We also conclude that positive initial curvature 
supports and negative curvature depresses the effectiveness of inflation.


\section{Hamilton-Jacobi equation for effective string action}  

Our starting point is the low energy effective action 
\begin{eqnarray} \label{3+1}
    I_{eff} = \frac{1}{2\lambda_{s}^{2}} \int d^{4}x \sqrt{-g} e^{-\phi} 
    [~^{4}{\mathcal{R}} + g^{\mu\nu} \partial_\mu \phi \partial_\nu \phi ] .
\end{eqnarray}
In the Arnowitt-Deser-Misner (ADM) formalism the dynamics is encoded in 
the Hamiltonian constraint \cite{saygili}
\beq \label{eseos}
    {\mathcal{H}} [ \pi^{ij},\pi^\phi,\gamma_{ij},\phi ]  \equiv
     \frac{e^\phi}{\sqrt\gamma} \ \left[\ \pi^{ij} \pi^{kl} \gamma_{ik} 
      \gamma_{jl} + \frac1 {2} {(\pi^\phi)}^2 + \pi \pi^\phi \ \right] - 
\nonumber\\[2ex]
    - \sqrt\gamma e^{-\phi}R
    - \sqrt\gamma e^{-\phi} \gamma^{ij}\partial_i \phi  \partial_j \phi + 
    2 \sqrt\gamma \Delta e^{-\phi} = 0 ,
\eeq 
where $R$ is the 3-dimensional scalar curvature of a spacelike 
hypersurface. 
The momentum constraints \cite{saygili}
\beq \label{iseos}
    {\mathcal{H}}_i[ \pi^{ij},\pi^\phi,\gamma_{ij},\phi ] \equiv  
    - 2(\gamma_{ik}\pi^{kj}),_j + \pi^{kl}\gamma_{kl,l} +
    \pi^\phi \phi_{,i} = 0 .
\eeq
state the reparametrization invariance on spatial hypersurfaces 
$x^{l} \rightarrow x^{l} + \xi^{l}$ (diffeomorphism invariance) 
\cite{halliwell}.

The equations of motion for $\phi$ and $\gamma_{ij}$ read
\cite{saygili}
\begin{equation} \label{fii.}
    \frac{1}{N} \left(\dot\phi-N^{i} \phi_{,i} \right) = 
    \frac{e^\phi}{\sqrt\gamma} \left(\pi^\phi + \pi \right) \, ,
\end{equation} 
\begin{equation} \label{gamma.}
    -2 K_{ij} \equiv
    \frac{1}{N} \left(\dot\gamma_{ij} - N_{i;j} - N_{j;i} \right) = 
    \frac{e^\phi}{\sqrt\gamma} \left(2\pi^{kl} \gamma_{ik} 
    \gamma_{jl} + \gamma_{ij}\pi^\phi \right) \, . 
\end{equation}
Here $K_{ij}$ is the 3-dimensional extrinsic curvature tensor (the 
trace of $K_{ij}$ is a generalization of the Hubble parameter) and 
$\pi=\gamma_{ij} \pi^{ij}$ is the trace of the gravitational momentum tensor.
Lapse function $N$ and shift vector $N^{i}$ describe the ADM 
3+1 decomposition of the spacetime.  
In the following we use a synchronous gauge, i.e. we put 
$N(t,x^{k})=1$ and  $N^{i}(t,x^{k})=0$. 

The solutions of equations for the momenta  can be determined 
from a particular solution of the
corresponding Hamilton-Jacobi equation \cite{saygili}
\beq \label{hje} 
    \frac{e^{\phi}}{\sqrt\gamma}\left[\frac{\delta S}{\delta\gamma_{ij}}
    \frac{\delta S}{\delta\gamma_{kl}} \gamma_{ik} \gamma_{jl} + 
    \frac1 2 \left(\frac{\delta S}{\delta\phi}\right)^2 + \gamma_{ij} 
    \frac{\delta S}{\delta\gamma_{ij}} \frac{\delta S}{\delta\phi}\right] 
\nonumber\\[2ex]
    -\sqrt\gamma e^{-\phi}R - \sqrt\gamma e^{-\phi} \gamma^{ij}
    \partial_i\phi \partial_j\phi + 2 \sqrt\gamma \Delta e^{-\phi} = 0, 
\eeq
as 
\begin{equation} \label{impulsid}
    \pi^\phi  = \frac{\delta S} {\delta \phi}, \,\qquad
    \pi^{ij} = \frac{\delta S} {\delta \gamma_{ij}}.
\end{equation} 
The Hamilton-Jacobi equation is obtained by replacing 
momenta (\ref{impulsid})
into the Hamiltonian constraint (\ref{eseos}).

After solving the Hamilton-Jacobi equation for $S$, we get an opportunity
to solve also  field equations (\ref{fii.}), (\ref{gamma.}), and 
consequently determine the evolution of 3-metric and dilaton. 
Although there is no hope to find an exact solution to 
the Hamilton-Jacobi equation,
it is possible to obtain approximate solutions.


\section{Long-wavelength approximation} 

We use the basic formalism of gradient expansion \cite{perry}
for investigating the Hamilton-Jacobi equation (\ref{hje}).
Functional $S[\gamma_{ij}(x),\phi(x)]$ is expanded 
in a  series of terms according to the number  of spatial gradients 
they contain:
\begin{equation} \label{S}
    S[\gamma_{ij}(x),\phi(x)] = S^{(0)} + S^{(2)} + S^{(4)} + \dots .
\end{equation}
Here $S^{(0)}$ contains no spatial gradients, $S^{(2)}$ contains two spatial 
gradients, and so on.
Solving the Hamilton-Jacobi equation order-by-order amounts to
requiring the Hamiltonian constraint to vanish at each order. 
We assume that each term in  expansion (\ref{S}) satisfy also the momentum 
constraint (\ref{iseos}).

The long-wavelength approximation is actually the assumption that the
characteristic comoving coordinate scale of spatial variations $L_{cm}$ 
(wavelength of inhomogeneities) is larger than the comoving Hubble radius 
$d_{cm}=(H\ a)^{-1}$, $d_{cm} << L_{cm}$. In terms of physical scale
($L_{ph}=a\ L_{cm}$), the same is written as $H^{-1} << L_{ph}$.
On scales less than $L_{ph}$ the time derivatives dominate over
spatial gradients and space is almost homogeneous.
Inside the homogeneous region the zeroth order three-metric and dilaton are 
time dependent, but approximately coordinate independent quantities.
During the pre-big bang superinflation, 
the Hubble radius $H^{-1}$  shrinks in time  
and therefore the efficiency of the gradient expansion
grows in time, as distinct from  the  case of the 
de Sitter inflation, where the Hubble radius remains approximately constant.

In the standard pre-big bang cosmology it is assumed that the
initial state was an approximately flat
spacetime. However, in the asymptotic past, 
it is natural to assume a generic (classical) initial state. 
Using the gradient expansion method,  
we can investigate a generic initial state perturbatively, 
i.e. in the second approximation we incorporate 
the spatial gradients  which represent the curvature and dilaton 
inhomogeneities.


\section{The zeroth order solution} 

In the zeroth order (long-wavelength approximation), 
the Hamilton-Jacobi equation  does not contain 
spatial derivatives and we can neglect the last three terms in 
equation (\ref{hje}) 
\begin{equation} \label{hje0}    
    \frac{e^{\phi}}{\sqrt\gamma}\left[\frac{\delta S^{(0)}}{\delta\gamma_{ij}}
    \frac{\delta S^{(0)}}{\delta\gamma_{kl}} \gamma_{ik} \gamma_{jl} + 
    \frac1 2 \left(\frac{\delta S^{(0)}}{\delta\phi}\right)^2 + 
    \gamma_{ij} \frac{\delta S^{(0)}}{\delta\gamma_{ij}} 
    \frac{\delta S^{(0)}}{\delta\phi}\right] 
    = 0 .
\end{equation}
Let its solution be in the form 
(cf \cite{saygili})
\begin{equation} \label{S0}
    S^{(0)} = \frac{4}{\sqrt{3}-1} \int e^{-\phi} \, H[\phi(t,x^{k})] 
              \sqrt\gamma d^3 x .
\end{equation} 
The integral is taken over invariant three-volume $\sqrt{\gamma}d^{3}x $ and 
the functional is therefore diffeomorphism invariant.
As required, $S^{(0)}$ contains no spatial gradients.
The prefactor is chosen so that $H$ corresponds to the usual Hubble 
parameter in the long-wavelength approximation (see Sect. 6).
Upon substituting  (\ref{S0}) into equation (\ref{hje0}) the 
Hamilton-Jacobi equation in the zeroth approximation reduces to 
a differential equation
\begin{equation} \label{shje}
   {\left(\frac{\partial H}{\partial \phi}\right)}^2 +H\; 
   {\left(\frac{\partial H}{\partial \phi}\right)} - {\frac{1}{2}}\;H^2 = 0 .
\end{equation}
Direct integration yields 
\begin{equation} \label{h}
    H(\phi) = H_{0} e^{\left({\frac{ \pm \sqrt3 - 1}{2}} \right) \phi} .
\end{equation} 
Here $H_{0}$ is an integration constant which will be specified later.
Since $H(\phi)$ is proportional to the usual Hubble parameter  
we choose the upper sign in the exponent, because
this is in agreement 
with general principles of the standard pre-big bang model, 
which states that in the asymptotic 
past the Universe was sufficiently flat i.e. $\phi \rightarrow -\infty \ 
\Leftrightarrow \  H(\phi) \rightarrow 0$. 

Taking into account expressions for canonical momenta (\ref{impulsid}) 
and for $H$ (\ref{h}),
field equations (\ref{fii.}), (\ref{gamma.}) in the 
zeroth order read
\begin{equation} \label{valjav}
    \dot\phi^{(0)}= {2 \sqrt{3} \over \sqrt{3} -1} H,
    \qquad
    \dot\gamma_{ij}^{(0)} = 2 H \gamma_{ij} .
\end{equation}
Equation for  the zeroth order dilaton can be directly integrated
\beq \label{fii}
    \phi^{(0)} (t) = -{\frac{2}{\sqrt{3}-1}} \, \ln \, \left[\sqrt{3} \, H_{0}
    \,(\,t_{0} - t\,) \right] , \qquad t < t_{0}.
\eeq
Here $t_{0}$ is an integration constant which, in general, depends on
spatial coordinates $x^{l}$. 
But these zeroth order inhomogeneities of the dilaton field can be
removed by a suitable choice of synchronous gauge \cite{lifsh} 
and in the following we take $t_0 = const$.
Parameter $t_{0}$ corresponds, in our interpretation, to the moment
when quantum and string
effects become significant (classical singularity) and roughly to the 
moment when superinflation ends:
$t_{0}=t_{singularity} \approx t_{f}$. 
From solution (\ref{fii}) we also get a restrictive condition for
$H_{0}$, namely, it should be positive, $H_{0} > 0$.
For latter convenience, let us take $H_{0} = 1/\sqrt{3}$,
then the full zeroth order solution reads
\beq \label{1phi} 
    \phi^{(0)} (t) = -{\frac{2}{\sqrt{3}-1}} \, \ln \,(\,t_{0} - t\,) ,
    \qquad t < t_{0} ,
\eeq
\beq \label{1gamma}
    \gamma_{ij}^{(0)}(t,x^{l}) = (\,t_{0} - t\,)^{-\frac{2}{\sqrt{3}}} \, 
     k_{ij}(x^{l}) , \qquad t < t_{0} , 
\eeq
where the seed metric $k_{ij}(x^{l})$ is an arbitrary function of
spatial variables alone. In the zeroth order solution it is assumed that 
it describes the fluctuations
which have wavelengths larger than the Hubble radius and
inside it the geometry is  approximately homogeneous and flat. 


\section{The second order solution} 

The second order Hamilton-Jacobi equation reads
\beq \label{sohje}
    {\mathcal{H}}^{(2)} &=& 2 \, H \gamma_{ij} \, 
    \frac{\delta S^{(2)}}{\delta \gamma_{ij}} + 
    \frac{2 \sqrt{3}}{\sqrt{3}-1} \, H 
    {\frac{\delta S^{(2)}}{\delta \phi}}
\nonumber\\[2ex]
    & & - \sqrt\gamma e^{-\phi} R -
    \sqrt\gamma e^{-\phi} \gamma^{ij} \partial_i \phi \partial_j \phi +
    2 \sqrt\gamma \Delta e^{-\phi} = 0 . 
\eeq
The generating functional is assumed to be
\be \label{2gf}
    S^{(2)} = \int d^{3}x {\sqrt{\gamma}} \, (~J(\phi)R + 
    K(\phi)\partial_{i}\phi\partial^{i}\phi~).
\ee
It contains 
terms up to the second order in spatial gradients and is diffeomorphism 
invariant.  We  neglect the term proportional to the third diffeomorphism 
invariant quantity $\partial_{i}\partial^{i}\phi$, since it is possible
absorb it into $K(\phi)$ \cite{perry}.
Upon calculating the variational derivatives of $S^{(2)}$,
substituting them into the second order Hamilton-Jacobi
equation  (\ref{sohje}) and grouping together 
the coefficients of $R$,  $\partial_{i}\phi \partial^{i}\phi$ and
$\partial_{i}\partial^{i}\phi$, we get  a system of equations
for $J(\phi)$ and $K(\phi)$
\be \label{1}
   H~J + \frac{2 \sqrt{3}}{\sqrt{3} - 1}
          ~{\frac{\partial J}{\partial\phi}}~H - e^{-\phi} = 0 \,,
\ee
\be \label{2}
   4~H~{\frac{\partial J}{\partial\phi}} + 
    \frac{4~\sqrt{3}}{\sqrt{3}-1}~H~K + 2~e^{-\phi} = 0 \,,
\ee
\be \label{3}
    -4~H~{\frac{\partial^{2}J}{\partial\phi^{2}}} + ~H~K     
    - \frac{2~\sqrt{3}}{\sqrt{3}-1}~H~{\frac{\partial K}{\partial\phi}}     
    + e^{-\phi} = 0.
\ee
Although there are three equations for two functions $J(\phi)$ and $K(\phi)$,
the system is not over-determined, since only two of them are independent.
From equations (\ref{1}) and (\ref{2}) we get particular solutions for 
$J(\phi)$ and $K(\phi)$, respectively
\be \label{J}
    J(\phi)=-\frac{\sqrt{3}(\sqrt{3}-1)}{4} e^{-(\frac{1+
             \sqrt{3}}{2}) \phi} , 
\ee
\be \label{K}
    K(\phi)=-\frac{3(\sqrt{3}-1)}{4} e^{-(\frac{1+\sqrt{3}}{2}) \phi} .
\ee
They satisfy also  the third equation (\ref{3}). 

We are considering the case when the first order dilaton field 
has only  time dependence, $t_0 = const$.
In this case the spatial derivates of the dilaton are absent from
field equations and we need only $J(\phi)$ for the following treatment.
We can write the field equations for the dilaton (\ref{fii.}) 
and for the  three-metric (\ref{gamma.}) up to  the second approximation
as follows 
\beq \label{f1}
    \dot{\phi} &=& \sqrt{3}(\sqrt{3}+1) \, H -  
   \frac{\sqrt{3}}{2} \, e^{\phi} \, J \, R(\gamma) ,
\\[3ex] \label{g1}
    {\dot{\gamma}}_{ij} &=& 2~H~\gamma_{ij}  
     -  {\frac{1}{2}}~e^{\phi}~J 
    \left[(\sqrt{3}-1)\gamma_{ij}R(\gamma) + 4R_{ij}(\gamma)~ \right]~.
\eeq
Introducing the gradient expansion explicitly and taking into account
expression (\ref{h}) for $H(\phi)$ we have
\beq \label{f2}
  \dot{\phi}^{(0)} + \dot{\phi}^{(2)}
&=& (\sqrt{3}+1)\,e^{(\frac{\sqrt{3}-1}{2})\phi^{(0)}} + 
             (t_{0} - t)^{-1}\,\phi^{(2)}  
\nonumber\\[2ex]           
         &+& \frac{3(\sqrt{3}-1)}{8}
             (t_{0} - t)\,R(\gamma)\,, 
\\[2ex] \label{g2}
  \dot{\gamma}_{ij}^{(0)} + \dot{\gamma}_{ij}^{(2)} 
&=& \frac{2}{\sqrt{3}}\,e^{(\frac{\sqrt{3}-1}{2})\phi^{(0)}}
                      \gamma_{ij} +  
  \frac{(\sqrt{3}-1)}{\sqrt{3}}\,(t_{0} - t)^{-1}\,\phi^{(2)}\, \gamma_{ij}
\nonumber\\[2ex]     
    &+&\frac{\sqrt{3}(\sqrt{3} - 1)}{8} (t_{0} - t)
     \left [ \left (\sqrt{3}-1 \right) \gamma_{ij} R(\gamma) +
     2R_{ij}(\gamma)~ \right].
\eeq
Solving  equation (\ref{f2})
we get for the dilaton up to the second order in spatial gradients 
\beq \label{f3}
  \phi(t,x^l) &=& \phi^{(0)}(t) + \phi^{(2)}(t,x^l) \equiv \phi^{(0)}(t) + 
                \delta\phi(t,x^l) 
\nonumber\\[1ex]      
       &=& -{\frac{2}{\sqrt{3}-1}} \, \ln \,(\,t_{0} - t\,) 
           + \delta\phi_{0}(x^l)\,(\,t_{0} - t\,)^{-1} 
\nonumber\\[1ex]
       &-&{\frac{3\sqrt{3}}{4(11+5\sqrt{3})}}
          (\,t_{0} - t\,)^{(2 + \frac{2}{\sqrt{3}})} ~ R(k) \,.   
\eeq
Here $\delta\phi_{0}(x)$ is a space dependent integration constant 
and it may be interpreted as an initial dilaton perturbation. 
In the asymptotic past the term containing this constant is decaying and 
we may omit it there. However, during the inflation this term is growing 
(corresponding dilaton inhomogeneities are not homogenized) 
and becomes important near the singularity.  
This is in agreement with numerical treatment carried out by Chiba 
\cite{chiba1}.
For the three-metric up to the second order in spatial gradients we get
\beq \label{3-m}
  \gamma_{ij} &=& \gamma_{ij}^{(0)}(t,x^l) + \gamma_{ij}^{(2)}(t,x^l) 
      \equiv  a^{2}(t) k_{ij}(x^l) + 
       a^{2}(t) \delta k_{ij}(t,x^l) \nonumber\\[2ex] 
      &=& (t_{0} - t)^{-\frac{2}{\sqrt{3}}}~\{ k_{ij}(x^{l})+ 
       \frac{\sqrt{3}-1}{\sqrt{3}}(t_{0} - t)^{-1} \delta\phi_{0}(x^l) 
      k_{ij} (x^l) 
       \\[2ex]      
      &-&   \frac{3(\sqrt{3}-1)}{4(\sqrt{3}+1)}(t_{0} - 
           t)^{2+\frac{2}{\sqrt{3}}}
      \left [{\frac{(\sqrt{3}+1)}{(5\sqrt{3}+11)}}~k_{ij}(x^l) R(k) 
      +  R_{ij}(k) \right]\},   \nonumber
\eeq
where $R_{ij}(k)$ and $R(k)$ are the Ricci tensor and the scalar curvature
of the 3-hypersurface calculated from the seed metric $k_{ij}(x^{l})$.
In the second order we have incorporated the curvature as a small 
perturbation and expression (\ref{3-m}) represents a non-linear 
evolution of the curvature and initial inhomogeneities.
In expression (\ref{3-m}) the term containing the spatial curvature becomes 
negligible during the superinflation as $t \rightarrow t_{0}$ and the 
initial spatial curvature and associated inhomogeneities will decay.
This is in agreement  with the standard  
result, which states, that spatial curvature becomes negligible 
during any kind of inflation. 
In the pre-big bang cosmology the similar conclusion is obtained in
\cite{vene1, buonanno1} and confirmed by numerical calculations 
in \cite{maharana}.

As we can see from expressions (\ref{f3}) and (\ref{3-m}) the curvature 
terms generated by the seed metric become important as we go backwards 
in time. 
Since the curvature term is growing fast, we must restrict 
the treatment here
with requirement that the second approximation (gradient terms) stays 
smaller than the zeroth order solution, $|S^{(0)}| > |S^{(2)}|$. 
This condition allows us to estimate the initial spatial curvature, 
and on the  other hand indicates the validity of the approximation. 
If the initial 3-curvature and inhomogeneities are arbitrarily large, 
the gradient expansion cannot be used for discussing the problem of initial 
conditions. However, it can be used for investigating the combined 
restrictions on initial inhomogeneities  and  duration of the
inflation.   


\section{The influence of initial inhomogeneities}

Let us consider the problem of fine-tuning of
initial conditions and determine the requirement for initial
curvature radius for getting sufficient amount of inflation. 
We adopt the procedure presented by Nambu and Taruya \cite{nambu}
in the context of the de Sitter inflation to the pre-big bang cosmology.  
At each point $x^{l}$ one can define a local Hubble parameter by
\be \label{H1}
    \bar{H}  =  \frac{\dot{\gamma}}{\gamma} = 
        \frac{1}{6} {\gamma^{ij}}{\dot{\gamma}_{ij}}.
\ee
Using  expression (\ref{g1}) for the three-metric one can find 
\beq \label{H}
    \bar{H} &=& \frac{1}{\sqrt{3}} \frac{1}{(t_{0} - t)} + 
    \frac{\sqrt{3}}{24}(4-\sqrt{3})(t_{0} - t)^{(1+\frac{2}{\sqrt{3}})}~R(k) 
\nonumber\\[2ex]    
    &=& H \left[~1+\frac{(4-\sqrt{3})}{24}~\frac{1}{H^{2}}~R(\gamma^{(0)})
         ~\right] .
\eeq
Here $H =  (\sqrt{3} (t_{0} - t))^{-1}$ is the  Hubble 
parameter for the zeroth order solution.  
In the zeroth order, the Hubble horizon $H^{-1}$ is proportional to 
the event horizon $d_{e}$ (to be precise, $ d_{e}=(1+ \sqrt{3})^{-1} H^{-1}$). 
In the second order the corresponding expression reads 
\beq
  \bar{d}_{ph} \equiv \bar{H}^{-1} =\frac{d_{ph}}{\left[ 1 
               + \frac{(4-\sqrt{3})}{24}\frac{1}{H^{2}} 
               R(\gamma^{(0)})\right]} \,,  
\eeq
where $d_{ph} \equiv H^{-1}$.
We see that the positive spatial curvature $R(\gamma^{(0)}) > 0$ 
reduces the horizon size compared with the zeroth order case, 
$\bar{d}_{ph} <  d_{ph}$, and 
therefore favours the inflation.
The negative spatial curvature $R(\gamma^{(0)}) < 0$ 
on the other hand increases 
the horizon size $\bar{d}_{ph} > d_{ph}$ and through that 
it is unfavorable for inflation.
The situation is contrary to the case of the de Sitter inflation 
\cite{nambu, wald}. 

Our conclusions are valid for small curvatures.
If the curvature takes the value 
$R^{cr}(\gamma^{(0)}) = -\left(\frac{24}{4-\sqrt{3}}\right) H^{2}$
the horizon size is infinite, $d_{ph} \rightarrow \infty$, but in this case 
the gradient expansion is long ago broken because the requirement that 
$|S^{(0)}| > |S^{(2)}|$ is not valid.  
  
Using the expression for $\bar{H}$, it is possible to calculate the 
number of e-folds of growth in scale factor during the superinflation 
including the curvature corrections
\beq  \label{N1}
    \bar{N} & = & \ln ~\left( \frac{\bar{a}(t_{f})}{\bar{a}(t_{i})} \right) = 
          \int\limits_{t_{i}}^{~\ t_{f}}~dt~\bar{H}
\nonumber\\[2ex]          
    &\approx & \ln~\left(~\frac{t_{0}-t_{i}}{t_{0}-
t_{f}}~\right)^{\frac{1}{\sqrt{3}}}
        - \left[ \frac{4-\sqrt{3}}{16(\sqrt{3}+1)}~    
        (t_{0} - t)^{2+\frac{2}{\sqrt{3}}}~R(k) \right]_{t_{i}}^{t_{f}}        
\nonumber\\[2ex]
     & \approx & N_{0} - p (t_{0} - t_{f})^{2}~R^{f}(\gamma^{(0)})
          + p (t_{0} - t_{i})^{2}~R^{i}(\gamma^{(0)}).
\eeq
Here $t_{i}$, $t_{f}$ are the onset and end time of superinflation,
$R^{i}(\gamma^{(0)})$, $R^{f}(\gamma^{(0)})$ are the initial and 
the final spatial curvatures  and $p$ is a positive numerical 
prefactor ($p\approx 1/20$).
$N_{0}$ is the e-folding for the zeroth order solution, i.e. for spatially
flat case (calculated from $H$). The second term in expansion (\ref{N1}) is 
suppressed if $t \rightarrow t_{0}$ and we can write the expression for
$\bar{N}$ as follows
\be \label{N2}
    \bar{N}  \approx  N_{0} + p  (t_{0} - t_{i})^{2}~R^{i}(\gamma^{(0)}) 
                    = N_{0} + \frac{p}{3} \frac{1}{H_{i}^{2}}\, 
R^{i}(\gamma^{0}).
\ee
The effect of initial curvature (inhomogeneity) is the same as considered
above.


Now we assume, that in the zeroth order the superinflation is long enough to 
satisfy the phenomenological constraints.  
This means that the comoving Hubble radius must decrease at least $e^{60}$
times during the inflation \cite{turnwein}
\beq
    Z = {\frac{d_{cm}^{i}}{d_{cm}^{f}}} = \left(\frac{t_{0} - t_{f}}{t_{0} - 
t_{i}}\right)
        ^{1+\frac{1}{\sqrt{3}}} > e^{60}.
\eeq
In the zeroth order, $N_{0}$ and $Z$ are related by $Z = e^{(\sqrt{3} + 
1)N_{0}}$.
From this requirement we get the following constraint for the 
Hubble horizon $H_{i}^{-1}$  at the onset moment $t_{i}$ 
of the superinflation(we assume that $H_{f}^{-1} \approx \lambda_{s}$)
\be \label{ti}
      H_{i}^{-1} =  e^{\sqrt{3}N_{0}} \,\lambda_{s} = 
e^{(\frac{\sqrt{3}}{\sqrt{3}+1}) \, \ln Z}
                    \,\lambda_{s} .
\ee
The second term in expression (\ref{N2}) has to be small with respect 
to $N_{0}$ and taking into account the expression (\ref{ti}), we get for 
initial curvature
\be \label{Ri}
    R^{i}(\gamma^{(0)}) < {\frac{3N_{0}}{p}} \, H_{i}^{2} < 
{\frac{3}{\sqrt{3}+1}} \, 
         {\frac{\ln Z}{p}} \, e^{-\left({\frac{2\sqrt{3}}{\sqrt{3}+1}}\right) \, 
\ln Z} \,
          \lambda_{s}^{-2} , 
\ee
and for initial curvature radius
\be \label{Ci}
    C^{i}_{curv} = \sqrt {\frac{6}{R^{i}(\gamma)}} = 
                   \sqrt {\frac{2 p(\sqrt{3} + 1)}{\ln Z}} \, 
                       e^{\left({\frac{\sqrt{3}}{\sqrt{3} + 1}}\right) \, ln Z}
                   \approx 0.07 H_{i}^{-1} \approx 10^{16} \lambda_{s}.
\ee
We see that the initial patch has to be extremely flat and also extremely
homogeneous because the characteristic size of inhomogeneities must be greater 
than the curvature radius $L^{i} > C^{i}_{curv}$. However, the 
characteristic size
of these inhomogeneities is only a small  part of the zeroth order horizon 
size (cf. Gasperini \cite{gasperini6}).  
This fact may be regarded as a fine-tuning of initial conditions. 
But it can also be regarded as a point of breakdown of
the gradient expansion method for considering the problem of initial
conditions. 
                              
\section{Summary}

In this paper we investigated the gradient expansion for the Hamilton-Jacobi 
equation derived from the low energy tree level effective string action. 
In the zeroth order we got cosmological pre-big bang  
solutions for the background with inhomogeneities
much larger than the Hubble radius (typical horizon scale).
The second order approximation includes the effect of spatial gradients.  
We find that  metric corrections die off during the superinflation
as $t \rightarrow t_{0}$, but  dilaton corrections are growing.
This means that initial classical inhomogeneities,
which ori\-ginate from spatial gradients of the seed metric,
are smoothed out during the superinflation, but not 
dilaton inhomogeneities.
Going backwards in time dilaton corrections become negligible, but
influence of initial classical inhomogeneities of the seed metric 
and the initial curvature are
growing.  Thus the adequacy  of the gradient expansion for investigating
the pre-big bang superinflation decreases
in the direction of the past as well as of the future (singularity).  
However, we can estimate the characteristic size of initial
inhomogeneities and conclude that the inflating domain must be large 
in string units but smaller than the initial horizon.    

\bigskip
\textbf{ACKNOWLEDGEMENTS}
\medskip 
We acknowledge useful comments of J.D.~Barrow and T.~Chiba on an earlier 
version of this paper.
This work was supported by the Estonian Science Foundation 
under grant No 3870.

\end{document}